\documentclass[aps,prl,twocolumn,groupedaddress,amsmath]{revtex4}

\usepackage{amsfonts}
\usepackage{amssymb}
\usepackage{amsmath}
\usepackage[dvips]{graphicx}
\usepackage{color}
\usepackage{amsfonts}
\usepackage{amssymb}
\usepackage{amsmath}
\usepackage[dvips]{graphicx}
\usepackage{color}

\usepackage{amsmath}
\usepackage{graphicx}
\usepackage{amsfonts}
\usepackage{amssymb}
\usepackage{hyperref}
\usepackage{color}
\usepackage{mathrsfs}
\usepackage{isomath}
\usepackage{amsmath}
\usepackage{amsthm}
\usepackage{epstopdf}
\usepackage{txfonts}

\begin{document}

\title{Sudden death and revival of Gaussian Einstein-Podolsky-Rosen steering
in noisy channels}
\author{Xiaowei Deng$^{1,2\dagger}$}
\author{Yang Liu$^{2,3\dagger}$}
\author{Meihong Wang$^{2,3}$}

\author{Xiaolong Su$^{2,3}$}
\email{suxl@sxu.edu.cn}

\author{Kunchi Peng$^{2,3}$}

\affiliation{$^1$Shenzhen Institute for Quantum Science and Engineering, Southern
University of Science and Technology Shenzhen 518055, China\\
$^2$State Key Laboratory of Quantum Optics and Quantum Optics Devices,
Institute of Opto-Electronics, Shanxi University, Taiyuan 030006, China\\
$^3$Collaborative Innovation Center of Extreme Optics, Shanxi
University, Taiyuan 030006, China\\
}

\begin{abstract}
Einstein-Podolsky-Rosen (EPR) steering is a useful resource for secure quantum information tasks. It is crucial to investigate the effect of inevitable loss and noise in quantum channels on EPR steering. We analyze and experimentally demonstrate the influence of purity of quantum states and excess noise on Gaussian EPR  steering by distributing a two-mode squeezed state through lossy and noisy channels, respectively. We show that the impurity of state never leads to sudden death of Gaussian EPR steering, but the noise in quantum channel can. Then we revive the disappeared Gaussian EPR steering by establishing a correlated noisy channel. Different from entanglement, the sudden death and revival of Gaussian EPR steering are directional. Our result confirms that EPR steering criteria proposed by Reid and I. Kogias et al. are equivalent in our case. The presented results pave way for asymmetric quantum information processing exploiting Gaussian EPR steering in noisy environment..
\end{abstract}

\maketitle

\section*{INTRODUCTION}

Nonlocality, which challenges our comprehension and intuition about the
nature, is a key and distinctive feature of quantum world. Three different
types of nonlocal correlations: Bell nonlocality \cite{Bell},
Einstein-Podolsky-Rosen (EPR) steering \cite%
{EPR1,EPR2,Review1,Review2,RMP2009} and entanglement \cite{Entanglement}
have opened an epoch of unrelenting exploration of quantum correlations
since they were introduced. The notion of EPR steering was introduced as the
phenomenon that one can remotely steer the (conditional) quantum state owned
by the other party through local measurements on half of an entangled state 
\cite{EPR1,EPR2}. According to the hierarchy of nonlocality, EPR steering
stands between entanglement and Bell nonlocality \cite{EPR from info}. The
existence of steering does not imply the violation of any Bell inequality,
while the violation of at least one Bell inequality immediately implies
steering in both directions which is referred as two-way steering \cite%
{OneWay1}. Remarkably, there is a situation that a quantum state may be
steerable from Alice to Bob, but not vice versa, which is called the one-way
steering \cite{OneWay1,OneWay2,DME,PRAQin,Caiyin,NC,PRL118,OneWay4,OneWay3}.
This intriguing feature comes from the intrinsically asymmetric with respect
to the two subsystems.

The one-way EPR steering has been observed for both continuous variable (CV) 
\cite{OneWay1,OneWay2,DME,PRAQin,Caiyin} and discrete variable (DV) systems 
\cite{NC,PRL118,OneWay4,OneWay3}. Gaussian states and Gaussian operations
play a central role in the analysis and implementation of CV quantum
technology \cite{Adesso,xin3}. When considering generally bipartite Gaussian
states and using Gaussian measurements, the steerability between the
submodes is referred as Gaussian steering. What's more interesting, it has
been shown that the direction of Gaussian EPR steering can be manipulated in
noisy environment \cite{PRAQin}. The temporal quantum steering and
spatio-temporal steering have also been investigated \cite{s1,PRA93,s2}.
Very recently, the EPR steering using hybrid CV and DV entangled state \cite%
{hybrid steering} and remote generation of Wigner-negativity through EPR
steering \cite{Wigner-neg} are demonstrated. As applications, EPR steering
has been identified as an essential resource for one-sided
device-independent quantum key distribution \cite{1sQKD2,1sQKD3,1sQKD4},
quantum secret sharing \cite{QSS3,QSS2}, secure quantum teleportation \cite%
{Tel1,Tel2,Tel3}, securing quantum networking tasks \cite{PRA99} and
subchannel discrimination \cite{Subchannel1,Subchannel2}.

Besides two-mode EPR steering, the multipartite EPR steering has also been
widely investigated since it has potential application in quantum network.
Genuine multipartite steering exists if it can be shown that a steering
nonlocality is necessarily shared among all observers \cite{xin10}. The
criterion for multipartite EPR steering and for genuine multipartite EPR
steering has been developed \cite{xin8,xin10}. Genuine high-order EPR
steering has also been experimentally demonstrated \cite{xin11}.
Multipartite Gaussian EPR steering in Greenberger-Horne-Zeilinger (GHZ)
state \cite{OneWay2,OCdeng} and a four-mode cluster state \cite{DME} have
been demonstrated. The experiments of multipartite Gaussian EPR steering
only demonstrated EPR steering in the case of bipartite splitting at
present, while genuine multipartite EPR steering for Gaussian states has not
been demonstrated. Recently, the distribution of multipartite steering in a
quantum network by separable states has been demonstrated \cite%
{xiangyu,wangmh}.

The performance of modern quantum communication sort of relies on the
interaction between quantum system and environment, which is generally
described by the theory of decoherence. Decoherence, caused by the loss and
noise encountered in quantum channel, can destroy fragile quantum properties
of quantum states, and thus the information carried by the quantum states
will be affected. Decoherence effects on entangled state has been well
studied \cite{ESD1,ESD2,esd3,esd4,pra83,esd5}. It has been shown that two
entangled qubits become completely separable in a finite-time under the
influence of vacuum which is the so called entanglement sudden death \cite%
{ESD1,ESD2}, and disentanglement occurs in Gaussian multipartite entangled
states when one mode is transmitted in a noisy channel \cite{NonM3}. It is
essential to recover entanglement when entanglement sudden death happens.\ A
number of methods to recover the destroyed entanglement have been
demonstrated, such as the non-Markovian environment \cite{NonM1,NonM2,NonM3}%
, weak measurement \cite{weakm} and feedback \cite{feedb}. Since EPR
steering is so intriguing and has enormous potential in quantum
communication applications, it is imperative and significant to investigate
the influence on Gaussian EPR steering made by decoherence. In 2015, the
decoherence of steering was theoretically investigated \cite{Josab15}.
Further more, it is unknown whether the methods used to recover entanglement
can be used to recover EPR steering when sudden death of EPR steering
happens.

Here, we experimentally investigate properties of Gaussian EPR steering for
a two-mode squeezed state (TMSS) when it is transmitted in lossy and noisy
quantum communication channels, respectively. Firstly, we investigate the
influence of purity of quantum states on Gaussian EPR steering. We show that
in a lossy channel, the impurity of the state leads to decrease of
steerabilities and two-way steering region, but it never leads to sudden
death of Gaussian EPR steering. Secondly, we find the Gaussian EPR steering
totally disappear at certain noise level in a noisy quantum channel, which
demonstrates sudden death of EPR steering. Thirdly, we successfully revive
the Gaussian EPR steering after its sudden death by establishing a
correlated noisy channel. We also confirm that the steering quantifier
proposed by I. Kogias et al. and the steering criterion proposed by M. Reid
are equivalent for a TMSS with Gaussian measurement. The presented results
provide useful references for applying Gaussian EPR steering in noisy
environment.

\section*{Experimental scheme}

\textbf{The principle}

\begin{figure*}[tbp]
\begin{center}
\includegraphics[width=120mm]{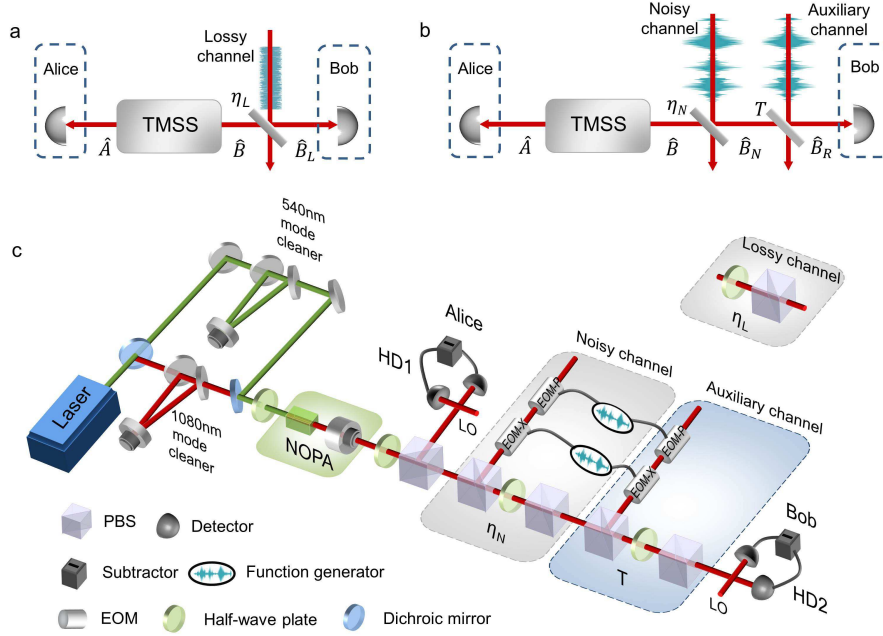}\vspace*{-0.3cm}
\end{center}
\caption{The principle of experiment. \textbf{a,} Schematic of EPR steering
in lossy channel. One mode of the two-mode squeezed state (TMSS) $\hat{A}$
is kept by Alice, the other mode of TMSS $\hat{B}$ is distributed to Bob
over a lossy channel which is experimentally mimicked by a beam-splitter,
the vacuum mode couples into system from the other input port of the
beam-splitter. The transmission efficiency of the lossy channel is $\protect%
\eta _{L}$. \textbf{b,} Schematic of disappearance and revival of EPR
steering. Mode $\hat{B}$ is distributed to Bob over a noisy channel with an
excess noise higher than the vacuum noise. The transmission efficiency of
the noisy channel is $\protect\eta _{N}$. An ancillary beam with noise
correlated to noisy channel couples with the transmitted mode on a
beam-splitter with transmission coefficient $T$. \textbf{c,} The
experimental setup. A pure -3 dB TMSS at the sideband frequency of 3 MHz is
generated by a nondegenerate optical parametric amplifier (NOPA). Alice and
Bob perform homodyne measurement on their states respectively. An ancillary
beam is used to revive the EPR steering. PBS, polarization beam splitter;
LO, local oscillator; EOM, electro-optic modulator; HD, homodyne.}
\end{figure*}

The state we use in the experiment is a deterministically prepared TMSS,
whose quantum correlations between quadratures are expressed by $\Delta
^{2}\left( \hat{x}_{A}+\hat{x}_{B}\right) =\Delta ^{2}\left( \hat{p}_{A}-%
\hat{p}_{B}\right) =2e^{-2r}$, where $\hat{x}=\hat{a}+\hat{a}^{\dagger }$
and $\hat{p}=\left( \hat{a}-\hat{a}^{\dagger }\right) /i$ are the amplitude
and phase quadratures of an optical mode respectively, $r$ is the squeezing
parameter ranging from 0 to infinite which correspond to no squeezing and
the ideal squeezing respectively. Under this definition, the noise of the
vacuum state is normalized to 1. All Gaussian properties of a TMSS can be
described by its covariance matrix%
\begin{equation}
\sigma _{AB}=\left( 
\begin{array}{cccc}
\alpha  & 0 & \gamma  & 0 \\ 
0 & \alpha  & 0 & -\gamma  \\ 
\gamma  & 0 & \beta  & 0 \\ 
0 & -\gamma  & 0 & \beta 
\end{array}%
\right) =\left( 
\begin{array}{cc}
\mathbf{A} & \mathbf{C} \\ 
\mathbf{C}^{T} & \mathbf{B}%
\end{array}%
\right) 
\end{equation}%
with $\sigma _{ij}=Cov\left( \hat{R}_{i},\hat{R}_{j}\right) =\frac{1}{2}%
\left\langle \hat{R}_{i}\hat{R}_{j}+\hat{R}_{j}\hat{R}_{i}\right\rangle
-\left\langle \hat{R}_{i}\right\rangle \left\langle \hat{R}_{j}\right\rangle 
$, $i,j=1,2,3,4$, where $\hat{R}=(\hat{x}_{A},\hat{p}_{A},\hat{x}_{B},\hat{p}%
_{B})^{T}$ is a vector composed by the amplitude and phase quadratures of
bipartite optical beams \cite{Adesso}. Thus the TMSS covariance matrix can
be partially expressed as (the cross correlations between different
quadratures of one mode are taken as $0$) 
\begin{equation}
\sigma =\left[ 
\begin{array}{cccc}
\bigtriangleup ^{2}\hat{x}_{A} & 0 & Cov\left( \hat{x}_{A},\hat{x}%
_{B}\right)  & 0 \\ 
0 & \bigtriangleup ^{2}\hat{p}_{A} & 0 & Cov\left( \hat{p}_{A},\hat{p}%
_{B}\right)  \\ 
Cov\left( \hat{x}_{A},\hat{x}_{B}\right)  & 0 & \bigtriangleup ^{2}\hat{x}%
_{B} & 0 \\ 
0 & Cov\left( \hat{p}_{A},\hat{p}_{B}\right)  & 0 & \bigtriangleup ^{2}\hat{p%
}_{B}%
\end{array}%
\right] .
\end{equation}%
For the theoretical covariance matrix elements of the TMSS we used in the
experiment, $\mathbf{A}=\alpha \mathbf{I}$, $\mathbf{B}=\beta \mathbf{I}$, $%
\mathbf{C}=\gamma \mathbf{Z}$, $\mathbf{I}$ and $\mathbf{Z}$ are the Pauli
matrices%
\begin{equation}
\mathbf{I}=\left( 
\begin{array}{cc}
1 & 0 \\ 
0 & 1%
\end{array}%
\right) ,\mathbf{Z}=\left( 
\begin{array}{cc}
1 & 0 \\ 
0 & -1%
\end{array}%
\right) 
\end{equation}%
and 
\begin{eqnarray}
\alpha  &=&\frac{V_{s}+V_{as}}{2}  \notag \\
\beta  &=&\frac{V_{s}+V_{as}}{2} \\
\gamma  &=&\frac{V_{s}-V_{as}}{2}  \notag
\end{eqnarray}%
in which $V_{s}=e^{-2r}$ and $V_{as}=e^{2r}+\delta $ are the squeezing and
anti-squeezing fluctuation of the initial squeezing state, respectively. The
existence of $\delta $ is caused by classical and uncorrelated noise in
quantum resource and leads to the impurity of quantum resource. To partially
reconstruct all relevant entries of its associated covariance matrix we
perform $6$ different measurements on the output optical modes. These
measurements include the amplitude and phase quadratures of the output
optical modes $\bigtriangleup ^{2}\hat{x}_{A},\bigtriangleup ^{2}\hat{p}%
_{A},\bigtriangleup ^{2}\hat{x}_{B},\bigtriangleup ^{2}\hat{p}_{B}$, and the
cross correlations $\Delta ^{2}\left( \hat{x}_{A}+\hat{x}_{B}\right) $ and $%
\Delta ^{2}\left( \hat{p}_{A}-\hat{p}_{B}\right) $. The covariance elements
are calculated via the identities \cite{recontructcm}%
\begin{align}
Cov\left( \hat{R}_{i},\hat{R}_{j}\right) & =\frac{1}{2}\left[ \Delta
^{2}\left( \hat{R}_{i}+\hat{R}_{j}\right) -\Delta ^{2}\hat{R}_{i}-\Delta ^{2}%
\hat{R}_{j}\right] ,  \notag \\
Cov\left( \hat{R}_{i},\hat{R}_{j}\right) & =-\frac{1}{2}\left[ \Delta
^{2}\left( \hat{R}_{i}-\hat{R}_{j}\right) -\Delta ^{2}\hat{R}_{i}-\Delta ^{2}%
\hat{R}_{j}\right] .
\end{align}%
As an example, the partially reconstructed covariance matrix of the
experimentally prepared pure TMSS is 
\begin{equation}
\sigma _{AB}=\left( 
\begin{array}{cccc}
1.26 & 0 & -0.79 & 0 \\ 
0 & 1.28 & 0 & 0.80 \\ 
-0.79 & 0 & 1.32 & 0 \\ 
0 & 0.80 & 0 & 1.28%
\end{array}%
\right) .
\end{equation}%
With these theoretical and experimentally partially reconstructed covariance
matrix, we can quantify the Gaussian EPR steering of quantum states.

The steerability of Bob by Alice ($A\rightarrow B$) for a two-mode Gaussian
state with Gaussian measurement can be quantified by \cite{criterion} 
\begin{equation}
\mathcal{G}^{A\rightarrow B}(\sigma _{AB})=\max \bigg\{0,\frac{1}{2}\ln 
\frac{\det \mathbf{A}}{\det \sigma _{AB}}\bigg\}.  \label{eqn:parameter}
\end{equation}%
The quantity $\mathcal{G}^{A\rightarrow B}$ vanishes iff the state described
by $\sigma _{AB}$ is nonsteerable by Gaussian measurements. The steerability
of Alice by Bob ($B\rightarrow A$), quantified by $\mathcal{G}^{B\rightarrow
A}$, can be obtained by replacing the role of A by B.

It is worth mentioning that the first criterion for the demonstration of the
EPR paradox was proposed by M. Reid in 1989 \cite{ReidPRA}, and it was
linked to the concept of steering lately \cite{PRA80}. For a two-mode
Gaussian state, the quantifier as given by Eq. (4) is equivalent to the
Reid's criterion which is based on Heisenberg uncertainty relation, i.e., a
state is steerable from Alice to Bob if the following relation is violated: 
\begin{equation}
V_{X_{B}|X_{A}}V_{P_{B}|P_{A}}\geq 1,
\end{equation}%
where $V_{X_{B}|X_{A}}$ and $V_{P_{B}|P_{A}}$ are the conditional variances
of Bob's measurements conditioned on Alice's results, and $%
V_{X_{B}|X_{A}}V_{P_{B}|P_{A}}=\det \sigma _{AB}/\det \mathbf{A}$. For
steering from Bob to Alice, the roles of A and B are swapped. This criterion
does not require the assumption of Gaussian states or Gaussian systems to be
valid.

The PPT criterion \cite{ppt} is applied to characterise the entanglement,
which is a sufficient and necessary condition for a Gaussian TMSS. The PPT
value can be determined by 
\begin{equation}
\sqrt{\frac{\Gamma -\sqrt{\Gamma ^{2}-4\det \sigma _{AB}}}{2}},
\end{equation}%
where $\Gamma =\det \mathbf{A}+\det \mathbf{B}-2\det \mathbf{C}$. When the
PPT value is less than $1$, the two modes are entangled.

At first, we investigate the effect of purity of quantum state on EPR
steering. The purity of a quantum state $\varrho $ is $\mu \left( \varrho
\right) =\mathbf{Tr}\varrho ^{2}$ \cite{Adesso}, which varies in the range
of $\frac{1}{D}\leq \mu \leq 1$, where $D=\dim \mathcal{H}$ for a given
Hilbert space $\mathcal{H}$. The minimum is reached by the totally random
mixture, the upper bound is saturated by pure states. In the limit of CV
systems ($D\rightarrow \infty $ ), the minimum purity tends to zero. The
purity of any N-mode Gaussian state is only function of the symplectic
eigenvalues $\upsilon _{i}$ of covariance matrix $\sigma$ \cite{purity2},
which is given by 
\begin{equation}
\mu \left( \varrho \right) =\frac{1}{\prod_{i}\upsilon _{i}}=\frac{1}{\sqrt{%
\det \sigma }}.
\end{equation}%
Regardless of the number of modes, the purity of a Gaussian state is fully
determined by the global symplectic invariant $\det\sigma $ alone \cite%
{purity2}.

\begin{figure}[tbp]
\begin{center}
\includegraphics[width=80mm]{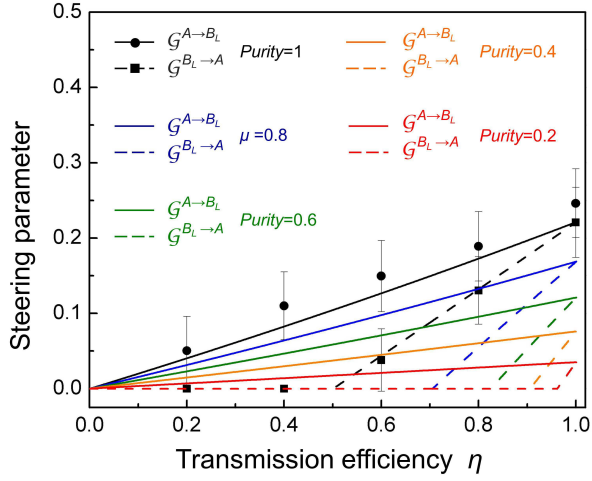}
\end{center}
\caption{The influence of purity of TMSS on the Gaussian EPR steering in a
lossy channel. \textbf{a,} Steering parameter quantified by Eq. (4). \textbf{%
b,} Steering certification according to Eq. (5). The region under the bound
of $1$ (black dotted line) indicates the existence of steerability.
Theoretical results for different purities of $1$, $0.8$, $0.6$, $0.4$ and $%
0.2$ are considered, which correspond to different $\protect\delta $ of $%
0,0.5,1.33,2.99$ and $7.97$, respectively. The experimental data (black
circles and squares) for the pure TMSS agree well with the theoretical
prediction (black solid and dash curves). Error bars of experimental data
represent $\pm 1$ standard deviation.}
\end{figure}

In practice, the loss and noise in the generation system will lead to
generation of impure quantum states \cite{esd5,PRA79}. For example, the
phonon noise in the three-color entangled state prepared by a non-degenerate
optical parametric oscillator (OPO) is a type of classical and uncorrelated
noise which leads to an impure three-color entangled state. Scattered light
by thermal phonons inside a second-order nonlinear crystal is the source of
additional phase noise observed in OPOs \cite{esd5,PRA79}. For our TMSS, the
relationship between purity $\mu \left( \varrho \right) $ and $\delta $ is 
\begin{equation}
\mu \left( \varrho \right) =\frac{1}{\sqrt{\det \sigma _{AB}}}=\frac{1}{%
V_{s}V_{as}}=\frac{e^{2r}}{e^{2r}+\delta }.
\end{equation}%

The scheme of the transmission of Gaussian EPR steering in a lossy channel
for pure and impure TMSSs is shown in Fig. 1a. Mode $\hat{B}$ of the TMSS is
distributed in a lossy channel, the mode after transmission is given by $%
\hat{B}_{L}=\sqrt{\eta _{L}}\hat{B}+\sqrt{1-\eta _{L}}\hat{v}$, where $\eta
_{L}$ and $\hat{v}$ represent the transmission efficiency of quantum channel
and the vacuum mode induced by loss into the quantum channel, respectively.
The partially reconstructed covariance matrix of the TMSS distributed over
lossy channel in the experiment we have%
\begin{eqnarray}
\alpha _{L} &=&\frac{V_{s}+V_{as}}{2}  \notag \\
\beta _{L} &=&\frac{\eta _{L}\left( V_{s}+V_{as}\right) }{2}+1-\eta _{L} \\
\gamma _{L} &=&\frac{\sqrt{\eta _{L}}\left( V_{s}-V_{as}\right) }{2}  \notag
\end{eqnarray}

Secondly, the transmission of Gaussian EPR steering in a noisy channel is
investigated, as shown in Fig. 1b, where the excess noise higher than the
vacuum noise exists. Mode $\hat{B}$ of the TMSS is distributed in the noisy
channel, the mode after transmission is given by $\hat{B}_{N}=\sqrt{\eta _{N}%
}\hat{B}+\sqrt{\left( 1-\eta _{N}\right) g}\hat{N}+\sqrt{1-\eta _{N}}\hat{v}$%
, where $\hat{N}$ and $g$ are the Gaussian noise in the channel with
transmission efficiency of $\eta _{N}$ and the magnitude of noise,
respectively. The partially reconstructed covariance matrix of the TMSS
distributed over noisy channel is 
\begin{eqnarray}
\alpha _{N} &=&\frac{V_{s}+V_{as}}{2}  \notag \\
\beta _{N} &=&\frac{\eta _{N}\left( V_{s}+V_{as}\right) }{2}+\left( 1-\eta
_{N}\right) \left( g+1\right)  \\
\gamma _{N} &=&\frac{\sqrt{\eta _{N}}\left( V_{s}-V_{as}\right) }{2}  \notag
\end{eqnarray}

Finally, we demonstrate the revival of Gaussian EPR steering by establishing
a correlated noisy channel, where an auxiliary channel carrying on the noise
correlated with that of the noisy channel is introduced. In today's
communication system, it is relevant to consider channels with correlated
noise (i.e. non-Markovian environment), because the time and space
correlated noise exist naturally \ \cite{correlated noise1,correlated
noise2,correlated noise3}. Furthermore, it has been experimentally found
that correlated noisy channel can be established by two bounded fibers \cite%
{bound fiber}, thus the method we use in experiment to built the
non-Markovian environment is applicable and practical significant. As shown
in Fig. 1b. The auxiliary noisy channel based on an ancillary coherent state 
$\hat{c}_{an}$, which is modulated to carry the noise information correlated
with that in the noisy quantum channel, saying $\hat{c}_{an}^{\prime }=\hat{c%
}_{an}+\sqrt{g_{an}}\hat{N}$ ($g_{an}$ is the magnitude of the correlated
noise and is experimentally adjustable). Thus there exists correlation
between the quantum system and environment. The noisy transmitted mode $\hat{%
B}_{N}$ couples with the auxiliary noisy channel on a revival beam-splitter
with transmission efficiency $T$. One of the output mode we need from the
revival beam-splitter is $\hat{B}_{R}=\sqrt{\eta _{N}T}\hat{B}+\sqrt{\left(
1-\eta _{N}\right) gT}\hat{N}-\sqrt{\left( 1-T\right) g_{an}}\hat{N}+\sqrt{%
\left( 1-\eta _{N}\right) T}\hat{v}-\sqrt{1-T}\hat{c}_{an}$, when the
adjustable $g_{an}$ and $T$ satisfy 
\begin{equation}
\frac{g}{g_{an}}=\frac{1-T}{\left( 1-\eta _{N}\right) T},  \label{condition}
\end{equation}%
it can be described as 
\begin{equation}
\hat{B}_{R}=\sqrt{\eta _{N}T}\hat{B}+\sqrt{\left( 1-\eta _{N}\right) T}\hat{v%
}-\sqrt{1-T}\hat{c}_{an}.
\end{equation}%
Thus the excess noise in mode $\hat{B}_{R}$ can be removed totally, and the
steerability between $\hat{B}_{R}$ and $\hat{A}$ can be revived to some
degree. After the revival operation, the covariance matrix of the revived
state is 
\begin{eqnarray}
\alpha _{R} &=&\frac{V_{s}+V_{as}}{2}  \notag \\
\beta _{R} &=&\frac{\eta _{N}T\left( V_{s}+V_{as}\right) }{2}+1-\eta _{N}T \\
\gamma _{R} &=&\frac{\sqrt{\eta _{N}T}\left( V_{s}-V_{as}\right) }{2}  \notag
\end{eqnarray}

\textbf{Experimental set-up}

The experimental set-up is shown in Fig. 1c. We use a non-degenerate optical
parametric amplifier (NOPA) to prepare the TMSS. NOPA is pumped by a
continuous-wave intracavity frequency-doubled Nd:YAP-LBO laser with two
output wavelengths at 540 nm and 1080 nm. The NOPA consists of an $\alpha $%
-cut type-II KTP crystal which is front face coated (work as the input
coupler with transmittance of 21.2\% and 0.04\% at 540 nm and 1080 nm,
respectively) and a concave mirror (work as the output coupler with
transmittance of 0.5\% and 12.5\% at 540 nm and 1080 nm, respectively). Via
the frequency down-conversion process of the pump field at 540 nm inside the
NOPA which operated at deamplification condition [the relative phase between
the pump laser and the injected signal equals to $\left( 2n+1\right) \pi $],
the TMSS at 1080 nm we want can be generated. Experimentally, the TMSS we
prepared is a nearly pure -3.0 dB state at sideband frequency of 3 MHz. 

The loss in the quantum channel is experimentally mimicked by a
beam-splitter, which is composed by a half-wave plate and a polarization
beam-splitters (PBS), the vacuum mode couples into system from the other
input port of the beam-splitter. In the noisy environment, to couple the
transmitted mode $\hat{B}$ and the noise-modulated coherent beam (auxiliary
noisy channel $\hat{c}_{an}^{\prime }$), the beam-splitter (revival
beam-splitter) is composed by a PBS, a half-wave plate and another PBS as
shown in Fig. 1c, where the transmission efficiency $\eta _{N}$ and $T$ are
adjusted by setting the half-wave plate. The correlated noises between the
quantum system and environment are generated by same signal generator
experimentally. We reconstruct the covariance matrix of the quantum state
with the homodyne detection systems, and quantify its steering properties.

\section*{RESULTS}

\textbf{The effect of purity on EPR steering}

\begin{figure*}[tbp]
\begin{center}
\includegraphics[width=145mm]{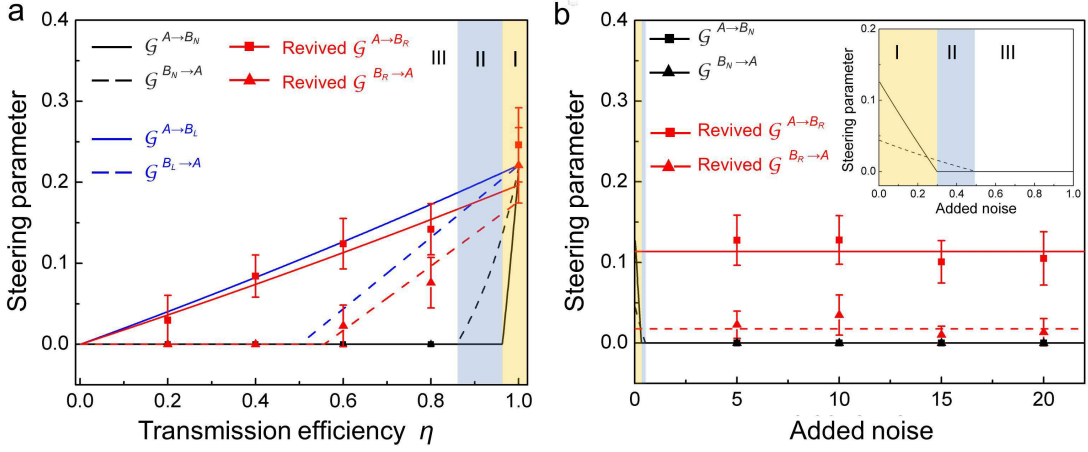}\vspace*{-0.3cm}
\end{center}
\caption{Disappearance and revival of EPR steering in noisy channel. \textbf{%
a,b} Steering parameters quantified by Eq. (4) and (5) in a noisy channel
where the variance of the excess noise is taken as $5$ times of vacuum
noise, respectively. Black dash and solid curves are the theoretical
predictions of $\mathcal{G}^{B_{N}\rightarrow A}$ and $\mathcal{G}%
^{A\rightarrow B_{N}}$, respectively. Region I, two-way steering; region II,
one-way steering; region III, no steering. Red dash and solid curves are the
theoretical predictions of $\mathcal{G}^{B_{R}\rightarrow A}$ and $\mathcal{G%
}^{A\rightarrow B_{R}}$ after steering revival, respectively. The Blue dash
and solid curves are the theoretical predictions of $\mathcal{G}%
^{B_{L}\rightarrow A}$ and $\mathcal{G}^{A\rightarrow B_{L}}$ in a pure
lossy but noiseless channel. \textbf{c,d} Steering parameters quantified by
Eqs. (4) and (5) in a transmission efficiency fixed ($\protect\eta _{N}=0.6$%
) noisy channel with different noise levels. Black dash and solid curves are
theoretical predictions of $\mathcal{G}^{B_{N}\rightarrow A}$ and $\mathcal{G%
}^{A\rightarrow B_{N}}$ as a function of the excess noise in the unit of
shot noise level. Red dash and solid curves are the theoretical predictions
of $\mathcal{G}^{B_{R}\rightarrow A}$ and $\mathcal{G}^{A\rightarrow B_{R}}$
after steering revival, respectively. The inset figure shows the detailed
theoretical predictions of $\mathcal{G}^{B_{R}\rightarrow A}$ and $\mathcal{G%
}^{A\rightarrow B_{R}}$ when the excess noise is relatively low. Error bars
of experimental data represent $\pm 1$ standard deviation and are obtained
based on the statistics of the measured data.}
\end{figure*}

Figure 2 shows the effect of purity of the TMSS on EPR steering in a lossy
channel, where the steerabilities quantified by two criteria given by Eqs.
(4) and (5) are shown in Fig. 2a and 2b, respectively. For a pure $-3$ dB
TMSS, one-way steering appears in a lossy channel, and the critical point of
one-way steering property occurs at $\eta _{L}=0.5$. For comparison, four
different purities of the TMSSs at $0.8$, $0.6$, $0.4$ and $0.2$ are
considered, which show that the maximum steerability and the phenomenon of
one-way EPR steering property happens earlier with the decreasing of purity.
This confirms that the impurity of the TMSS leads to decrease of
steerabilities and two-way steering region. However, the steerability from
Alice to Bob never disappear until the purity tends to zero, which means
that the impurity of the TMSS does not lead to sudden death of Gaussian EPR
steering. The results in Fig. 2a and 2b confirm that two criteria given by
Eqs. (4) and (5) are equivalent for a TMSS with Gaussian measurement in a
lossy channel.

\begin{figure*}[tbp]
\begin{center}
\includegraphics[width=180mm]{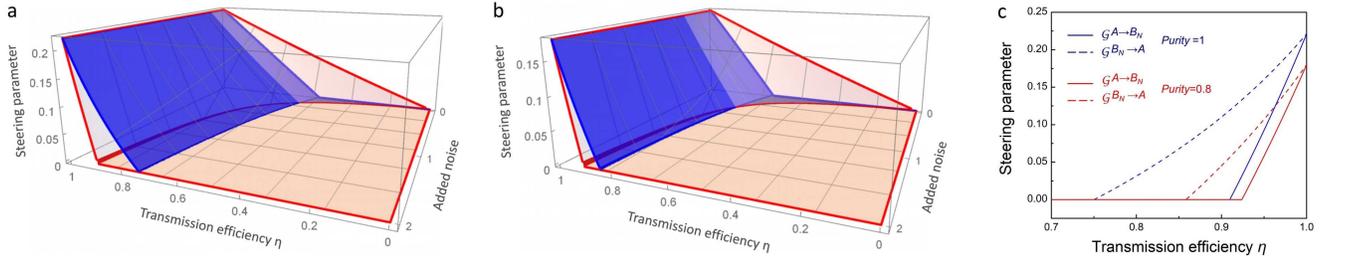}
\end{center}
\caption{Results for the effect of purity of the TMSS on EPR steering in a
noisy channel. \textbf{a, b,} The dependence of Gaussian EPR steering on
transmission efficiency and excess noise with initial pure and impure
(purity is 0.8) TMSS, respectively. The steerabilities of $\mathcal{G}%
^{A\rightarrow B_{N}}$ and $\mathcal{G}^{B_{N}\rightarrow A}$ are given by
the light red and blue colors, respectively. \textbf{c,} The effect of noise
on EPR steering of the TMSS with purities of 1 [blue curves] and 0.8 [red
curves], respectively. The excess noise is taken as $2$ times of vacuum
noise.}
\end{figure*}

\begin{figure}[tbp]
\begin{center}
\includegraphics[width=56mm]{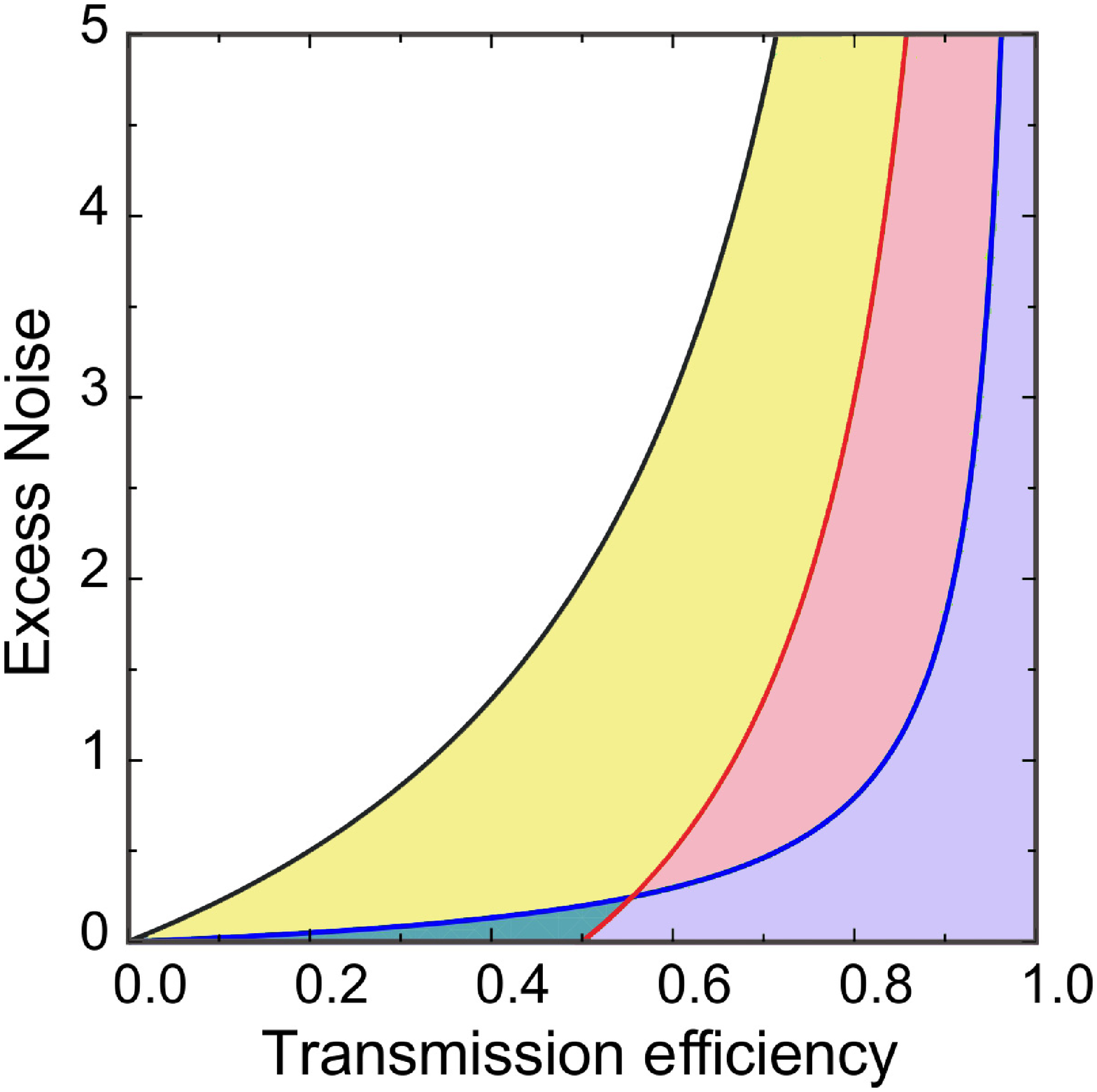}
\end{center}
\caption{The region plot parameterized by transmission efficiency and excess
noise. The region below the black curve, which is the bound of entanglement
sudden death, represents the existence of the entanglement. The region below
the blue and red curves, which are the bound of the disappearance of $%
\mathcal{G}^{A\rightarrow B}$ and $\mathcal{G}^{B\rightarrow A}$
respectively, represents the existence of EPR steering. }
\end{figure}

In the experiment, to determine the error bar of the experimental results,
we measure three covariance matrices for each quantum state and obtain the
corresponding EPR steering parameters. By calculating the mean value and the
standard deviation of these three EPR steering parameters, error bars are
obtained. The errors of the experimental data mainly come from the phase
fluctuation of the phase locking system, and we take much efforts to
suppress phase fluctuation of each phase locking system to be $\theta
\approx 1.5^{\circ }$ by adjusting locking parameters.

\textbf{The effect of noise on EPR steering}

As shown in Fig. 3a and 3b, sudden death of Gaussian EPR steering of the
TMSS in a noisy channel is observed, where the variance of the excess noise
is taken as $5$ times of vacuum noise. Specifically, in region I ( $%
0.96<\eta _{N}\leq 1$), the two-way steering exists, both Alice and Bob can
steer with each other. With the decreasing of $\eta _{N}$, when $0.86<\eta
_{N}\leq 0.96$ (labeled as region II), one-way steering appears, i.e. Bob
can steer Alice's state while Alice cannot steer Bob's state. More
seriously, when $\eta _{N}\leq 0.86$, the steerability between Alice and Bob
totally disappeared (labeled as region III), i.e. the sudden death of the
Gaussian EPR steering happens.

We also demonstrate sudden death of Gaussian EPR steering with different
excess noise levels (noise levels are taken as $5,10,15,20$ times of vacuum
noise) at fixed transmission efficiency ($\eta _{N}=0.6$) in Fig. 3c and 3d.
We can see the two-way steering (labeled as region I) and/or one-way
steering (labeled as region II) could still exist when the excess noise is
relatively low. When the variance of the excess noise is higher than 0.47
(labeled as region III), the sudden death of the Gaussian EPR steering
happens. The results in Fig. 3 confirm that two criteria given by Eqs. (4)
and (5) are equivalent for the sudden death and revival of Gaussian EPR
steering of a TMSS with Gaussian measurement.

The effect of purity of initial TMSS on EPR steering in a noisy environment
is also investigated as shown in Fig. 4. Figure 4a and 4b are the
three-dimensional theoretical results of Gaussian EPR steering properties of
initial TMSS with different purities of $1$ and $0.8$, respectively. It is
obvious that the direction of EPR steering is changed with different noise
levels, which is the same with the result in Ref. \cite{PRAQin}. We can see
that the blue surface ($\mathcal{G}^{B_{N}\rightarrow A}$) and light red
surface ($\mathcal{G}^{A\rightarrow B_{N}}$) have crossover, which happens
with excess noise of $0.24$ and $0.49$ for the pure and impure TMSSs,
respectively. Figure 4c shows the effect of purity of the TMSS on EPR
steering in a noisy channel with excess noise of $2$ times of vacuum noise.
It's obvious that the sudden death of EPR steering for the impure state
happens earlier and the maximum steerability is smaller than that of the
pure state.

\noindent\ \ \textbf{The revival of EPR steering}

To revive the disappeared EPR steering, we fix $T=0.9$ for the revival
beam-splitter and adjust the magnitude of correlated noise $g_{an}$. The
parameters $g/g_{an}$ are set to $0.14,0.18,0.28,$ and $0.56$ for different
quantum channel efficiencies of $0.2,0.4,0.6,$ and $0.8$, respectively,
according to Eq. (8). After the revival operation, the disappeared Gaussian
EPR steering is revived as shown in Fig. 3a and 3b. Because the excess noise
in mode $\hat{B}_{R}$ is removed totally in the revival operation, the
Gaussian EPR steering is revived even at the noise level of $20$ times of
vacuum noise , as shown in Fig. 3c and 3d. Generally, $T$ is better to be
close to $1$, because it results in adding an extra linear loss on the
revived mode, this is why the steerabilities between mode $\hat{B}_{R}$ and $%
\hat{A}$ are not good as that between mode $\hat{B}_{L}$ and $\hat{A}$ [the
pure lossy transmitted regime, presented by blue dash and solid curves in
Fig. 3a].
\begin{figure}[tbp]
\begin{center}
\includegraphics[width=80mm]{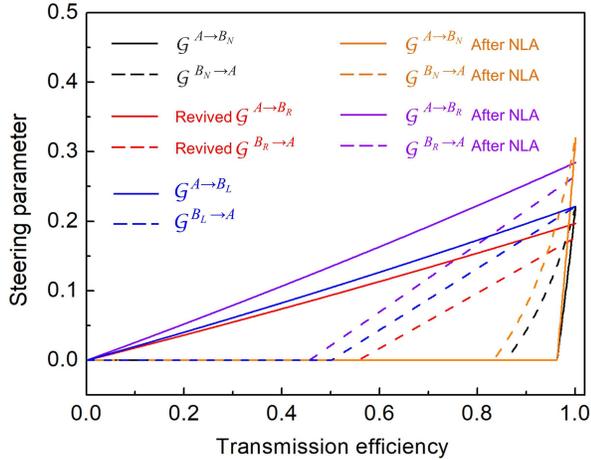}
\end{center}
\caption{Results for distillation of Gaussian steering with a correlated
noisy channel and measurement-based NLA. Black solid and dash curves are the
theoretical predicted steerabilities in a noisy channel with excess noise 5
times of vacuum noise. Red solid and dash curves are the predicted
steerabilities after revival of Gaussian steering with a correlated noisy
channel. The orange and purple curves are the predicted steerabilities when
the measurement-based NLA based on Bob's measurement results
with gain of 1.2 is applied before and after the revival of Gaussian
steering, respectively. The Blue solid and dash curves are the predicted
steerabilities in a pure lossy but noiseless channel. }
\end{figure}

\section*{DISCUSSION}

Note that the noise in quantum channel and the noise in initial TMSS have
different influences on EPR steerability. With the decreasing of purity of
initial TMSS, the maximum steerability decreases and the phenomenon of
one-way EPR steering property happens earlier than that of\ the pure state
(between $0.5\leq \eta \leq 1$), but the death of the EPR steering never
happens. While with the increasing of noise in the quantum channel, the
two-way EPR steering firstly declines to one-way EPR steering, and the
sudden death of the Gaussian EPR steering finally occurs. Thus we can
distinguish where the excess noise exists (whether in the communication
channel or in the initial TMSS) by the Gaussian EPR steerability easily.

The sudden death of Gaussian EPR steering is different from that of the
Gaussian entanglement, which is characterised by the positive partial
transposition (PPT) criterion \cite{ppt} (see Methods for details). Figure 5
shows the entanglement and steering region parameterized by transmission
efficiency and excess noise, where the purple region represents the two-way
steering region, the cyan region represents the one-way steering region of $%
\mathcal{G}^{A\rightarrow B}$, the light red region represents the one-way
steering region of $\mathcal{G}^{B\rightarrow A}$, and the yellow region
represents the entanglement without steering region, respectively. Firstly,
the direction of the EPR steering is changed with the increase of excess
noise which is not the case in entanglement sudden death. The sudden death
of EPR steering and entanglement happens when there is excess noise in the
quantum channel. For a pure $-3$ dB TMSS, the direction of EPR steering is
changed when the excess noise is 0.25 times of vacuum noise. This result
comes from the intrinsic asymmetric character of EPR steering. Secondly, it
is obvious that the sudden death of Gaussian EPR steering happens earlier
than that of the entanglement with same excess noise level. Thirdly,
different from entanglement, the sudden death of Gaussian EPR steering in a
noisy channel is directional.

Recovering EPR steering with correlated noisy channel can only remove the
effect of excess noise in the transmission channel, the effect of loss in
the transmission channel cannot be removed. Some other technologies such as
noiseless linear amplification can be used to eliminate the effect of loss 
\cite{linear amp1,linear amp2,linear amp3,linear amp4}. Recently, it has
been shown that the measurement-based noiseless linear amplification can be
used to distill Gaussian EPR steering in both lossy and noisy channels,
which provides another feasible way to recover Gaussian EPR steering in
quantum channel \cite{arxivLiu}.

Comparing the present work of EPR steering revival with a correlated noisy
channel and distillation of EPR steering with measurement-based NLA, the
differences are as following. Firstly, the steering revival with a
correlated noisy channel can only remove the effect of excess noise in the
quantum channel, the effect of loss in the transmitted channel cannot be
removed. But the distillation of EPR steering works for both lossy and noisy
channels. Secondly, the revival of Gaussian EPR steering with a correlated
noisy channel can remove the excess noise in the quantum channel totally,
but distillation of EPR steering with measurement-based NLA cannot remove
the excess noise in all range of transmission efficiency [orange curves in
Fig. 6]. Thirdly, the revived steerability with a correlated noisy channel
cannot exceed the initial steerability (that in a pure lossy channel), but
the distilled steerability can exceed the initial steerability.

Combining the methods of correlated noisy channel and the NLA may provide an
option to distill Gaussian EPR steering in lossy and noisy environment. As
shown in Fig. 6, if the measurement-based NLA based on Bob's
measurement results with gain of 1.2 is applied after the revival of
Gaussian EPR steering with a correlated noisy channel, the revived
steerability could be enhanced by using the measurement-based NLA. In this
case, the revival of steering with a correlated noisy channel removes the
effect of excess noise in quantum channel. Followed by the measurement-based
NLA, the effect of loss in quantum channel can be eliminated and the
distillation of Gaussian steering can be obtained.

Our work demonstrates evolution of bipartite Gaussian EPR steering of a TMSS
transmitted in lossy and noisy quantum communication channels. We show the
noise in quantum channel and the noise in the initial state result in
different influences on EPR steerability. The impurity of the state only
decrease the steerabilities and two-way steering region, but never leads to
sudden death of EPR steering. While the excess noise in the quantum channel
leads to sudden death of Gaussian EPR steering. We successfully revive EPR
steering after the sudden death occurs in noisy channel by establishing a
correlated noisy channel. The presented results provide useful references
for understanding the steering properties of quantum state which interacts
with different environments, also have potential applications in asymmetric
quantum information processing exploiting EPR steering as a valuable
resource.

\textbf{ACKNOWLEDGMENTS}

This research was supported by the NSFC (Grant Nos. 11904160, 11834010 and
62005149), the program of Youth Sanjin Scholar, National Key R\&D Program of
China (Grant No. 2016YFA0301402), and the Fund for Shanxi \textquotedblleft
1331 Project\textquotedblright\ Key Subjects Construction.

X. D. and Y. L. contributed equally to this work.

\end{document}